\documentclass{optica-article}

\journal{opticajournal} 

\articletype{Research Article}
\usepackage{float}
\usepackage{lineno}
\usepackage{amsmath}
\usepackage{soul, color, xcolor}
\usepackage{svg}
\usepackage{dcolumn}
\usepackage{multirow}
\usepackage{amsfonts}
\usepackage{booktabs}
\usepackage{graphicx}
\usepackage{comment}

\begin{document}

\title{Length-dependent SWIR upconversion spectral response of noncritically phase-matched KTP crystals}

\author{Xiao-Hua Wang,\authormark{1,2,3,*} Chang-Hao Min,\authormark{1,2,3,*} Zhao-Qi-Zhi Han,\authormark{1,2,3} Qi-Yu Chen,\authormark{1,2,3} Jin-Peng Li,\authormark{1,2,3} Bo-Wen Liu,\authormark{1,2,3} Yin-Hai Li,\authormark{1,2,3,4} Zhi-Yuan Zhou,\authormark{1,2,3,4,\dag} and Bao-Sen Shi\authormark{1,2,3,\ddag}}

\address{\authormark{1}Laboratory of Quantum Information, University of Science and Technology of China, Hefei 230026, China\\
\authormark{2}CAS Center for Excellence in Quantum Information and Quantum Physics, University of Science and Technology of China, Hefei 230026, China\\
\authormark{3}Anhui Province Key Laboratory of Quantum Network, University of Science and Technology of China, Hefei 230026, China\\
\authormark{4}Anhui Kunteng Quantum Technology Co. Ltd., Hefei 231115, China\\

\authormark{*}These authors contributed equally to this work.\\
}

\email{\authormark{\dag}zyzhouphy@ustc.edu.cn}
\email{\authormark{\ddag}drshi@ustc.edu.cn}



\begin{abstract*} 
Noncritically phase-matched (NCPM) KTP crystals support large-aperture bulk operation, avoid spatial walk-off, and relax angular-alignment requirements, making them attractive for short-wave infrared upconversion detection. To guide crystal selection for different detection requirements, we quantitatively characterize how crystal length affects the coverage and profile of their external upconversion spectral responses. A calibrated Czerny--Turner monochromator is used to measure the responses of 0.5, 1.0, and 2.0 mm crystals and compare them with theoretical quantum-efficiency spectra calculated from the phase-matching model. With increasing crystal length, the response evolves from a broad profile with weak peak separation to a more distinct double-peak structure, accompanied by reduced coverage bandwidth. A representative pump-power measurement further yields the system-level external quantum-efficiency slope. These results clarify the trade-off between spectral coverage and wavelength selectivity for different crystal lengths and can be used to choose the crystal length for SWIR upconversion detection systems.
\end{abstract*}

\section{Introduction}

Short-wave infrared (SWIR) spectral detection is important for infrared spectroscopy, imaging, free-space optical sensing, and related weak-light measurement scenarios. The 1--2.5~$\mu\mathrm{m}$ region contains useful atmospheric transmission windows and offers reduced scattering compared with visible wavelengths \cite{wilson2015review, carr2018shortwave, zhao2020shortwave, he2018crucial}. However, direct SWIR detection still faces practical limitations. InGaAs-based detectors are widely used in this spectral region, but their cost, dark noise, array scalability, and long-wavelength response remain limiting factors, especially for weak-light or broadband measurements \cite{rogalski2003infrared, zhu2024review}. Detectors with extended infrared response or higher sensitivity often require cooling or more complex system architectures \cite{zhang2018iga}. These limitations motivate frequency-upconversion detection, in which infrared photons are converted to shorter wavelengths by nonlinear optical mixing and then detected by mature silicon-based detectors.

Infrared upconversion detection is commonly implemented through sum-frequency generation (SFG), where an infrared signal and a pump beam generate a shorter-wavelength field in a second-order nonlinear medium. The response of this process is determined by three-wave mixing and phase matching \cite{boyd1968parametric, armstrong1962interactions}, while quasi-phase matching provides an effective route to efficient conversion in periodically poled media \cite{fejer1992quasi}. Upconversion spectrometers have been demonstrated in several forms, including waveguide-based systems for sensitive telecom-band measurements \cite{zhang2008waveguide, ma2009experimental}, tunable upconversion detectors and spectrometers \cite{kuo2013spectral, slattery2013tunable, shentu2013ultralow}, and more recent systems for mid-infrared detection, broadband spectroscopy, high-speed spectral acquisition, low-noise measurement, and quantum infrared spectroscopy \cite{johnson2012mid, hu2012high, dam2012room, wolf2017upconversion, barh2017ultra, rodrigo2021room, chen2023low, hashimoto2023upconversion, tashima2024ultra}. These studies show that nonlinear upconversion is a useful route for infrared spectroscopy. In an upconversion spectrometer, the conversion module determines the phase-matching bandwidth, response shape, and pump-power dependence, which in turn affect the usable spectral range and spectral calibration. Therefore, the spectral properties of the nonlinear conversion medium must be characterized quantitatively when it is used as a broadband upconversion module.

Many high-efficiency upconversion spectrometers employ quasi-phase-matched waveguides or periodically poled crystals. Although these devices provide high conversion efficiency, their small mode areas and limited apertures can restrict free-space collection and broadband spatial coupling. Bulk nonlinear crystals offer larger apertures and are therefore attractive for free-space architectures, but conventional critical phase matching generally requires angular tuning and may introduce angular sensitivity and spatial walk-off. Noncritical phase matching (NCPM) avoids spatial walk-off and relaxes the angular-alignment requirement, making bulk NCPM crystals well suited for free-space SWIR upconversion detection. Broadband SWIR upconversion imaging has recently been demonstrated using a 0.5~mm-long bulk NCPM KTP crystal with a $6~\mathrm{mm}\times7~\mathrm{mm}$ aperture over the 1.3--2.2~$\mu\mathrm{m}$ range \cite{wang2026short}. However, the effect of crystal length on the external upconversion spectral response has not been systematically quantified.

To address this gap, we use a calibrated Czerny--Turner (C--T) monochromator to characterize NCPM KTP crystals with lengths of 0.5, 1.0, and 2.0~mm. The measured responses are compared with theoretical quantum-efficiency spectra calculated using the NCPM phase-matching model, and their length-dependent evolution is quantified through the coverage bandwidth, peak ratio, and valley depth. We also determine the system-level external quantum-efficiency slope from a representative pump-power measurement in the low-pump-power region. The results provide guidance for selecting the crystal length and operating conditions of NCPM KTP conversion modules for SWIR upconversion spectrometers and detection systems.

\section{Principle and experimental methods}
\subsection{Wavelength selection and calibration using a C--T monochromator}
\label{subsec:CTMODEL}
A C--T monochromator was used to select a calibrated and tunable narrowband SWIR input for the subsequent measurements. The monochromator used a 600 lines/mm grating with an effective illuminated width of approximately $30~\mathrm{mm}$ and a focal length of $152.4~\mathrm{mm}$. In the monochromator, the diffraction grating and output aperture selected the desired SWIR wavelength from the broadband input light. An adjustable aperture was used for wavelength calibration and for the measurements of the upconversion response, whereas a 10 $\mu\mathrm{m}$ slit was used for the 1550 nm spectral-resolution verification. By scanning the grating angle, the selected signal wavelength could be tuned continuously over the measurement range.

The wavelength selected by the C--T monochromator is governed by the grating equation
\begin{equation}
d(\sin\alpha+\sin\beta)=m\lambda ,
\end{equation}
where $d$ is the grating period, $\alpha$ and $\beta$ are the incident and diffraction angles, $m$ is the diffraction order, and $\lambda$ is the selected wavelength. In the present configuration, the output direction was fixed and the grating angle was scanned by a motorized rotation stage. Therefore, the selected wavelength can be described as a function of the motor step number. The detailed form of this step--wavelength relation is given in the Supplementary Material.

The wavelength calibration was performed using the equivalence between first-order infrared diffraction and second-order visible diffraction at the same grating angle and output direction. Under this condition,
\begin{equation}
\label{eq:4}
\lambda_{\mathrm{IR}}=2\lambda_{\mathrm{vis}} .
\end{equation}
The second-order visible diffraction selected by this aperture was coupled into a spectrometer operating in the visible wavelength range, and its center wavelength was measured at each motor position. The corresponding first-order SWIR wavelength was then obtained from Eq.~(\ref{eq:4}). Under this adjustable-aperture configuration, the selected spectral width was approximately $0.4~\mathrm{nm}$, as shown in the Supplementary Material. Fitting these calibrated wavelength points to the grating relation above yielded the step--wavelength mapping.

In addition, a 1550 nm quasi-single-frequency laser was used to compare the spectral widths before and after upconversion. For this verification, the output aperture was replaced by a 10 $\mu\mathrm{m}$ slit. Under this slit-based configuration, the estimated spectral resolution is approximately $0.1~\mathrm{nm}$, mainly limited by the slit rather than by the grating diffraction limit. The wavelength increment corresponding to one motor step is about $0.0128~\mathrm{nm/step}$.

\subsection{Upconversion response model of NCPM KTP crystals}
\label{subsec:ktp_response_model}

The spectral response of the NCPM KTP crystals was calculated from the phase-matching condition of sum-frequency generation (SFG). In the present configuration, a narrowband SWIR signal at wavelength $\lambda_\mathrm{s}$ is upconverted by a fixed pump at wavelength $\lambda_\mathrm{p}$. The generated upconverted wavelength $\lambda_\mathrm{up}$ is determined by energy conservation,
\begin{equation}
    \frac{1}{\lambda_\mathrm{up}}
    =
    \frac{1}{\lambda_\mathrm{s}}
    +
    \frac{1}{\lambda_\mathrm{p}} .
    \label{eq:sfg_wavelength}
\end{equation}

For the type-II NCPM interaction used here, $n_\mathrm{s}$, $n_\mathrm{p}$, and $n_\mathrm{up}$ denote the refractive indices corresponding to the experimentally used signal, pump, and upconverted polarizations, respectively. The wave-vector mismatch is written as
\begin{equation}
    \Delta k(\lambda_\mathrm{s})
    =
    k_\mathrm{up}
    -
    k_\mathrm{s}
    -
    k_\mathrm{p}
    =
    2\pi
    \left(
    \frac{n_\mathrm{up}}{\lambda_\mathrm{up}}
    -
    \frac{n_\mathrm{s}}{\lambda_\mathrm{s}}
    -
    \frac{n_\mathrm{p}}{\lambda_\mathrm{p}}
    \right) .
    \label{eq:phase_mismatch}
\end{equation}
The refractive indices were calculated from the Sellmeier equations of KTP. The pump wavelength and crystal temperature were fixed in the calculation, and the signal wavelength was scanned over the measured SWIR range.

Under the low-conversion and undepleted-pump approximation, the wavelength-dependent upconversion quantum efficiency is governed mainly by the longitudinal phase-matching function. For a crystal with length $L$, the relative spectral response can be expressed as
\begin{equation}
    \eta(\lambda_\mathrm{s},L)
    \propto
    \frac{L^2}
    {n_\mathrm{s} n_\mathrm{p} n_\mathrm{up}
    \lambda_\mathrm{s}\lambda_\mathrm{up}}
    \,
    \mathrm{sinc}^2
    \left(
    \frac{\Delta k(\lambda_\mathrm{s})L}{2}
    \right),
    \label{eq:relative_efficiency}
\end{equation}
where $\mathrm{sinc}(x)=\sin(x)/x$. In this relative model, constants that do not vary with $\lambda_\mathrm{s}$, such as the effective nonlinear coefficient, pump power, and common focusing factors, are omitted. The calculated spectra were finally normalized to their respective maxima.

Eq.~\eqref{eq:relative_efficiency} contains two wavelength-dependent contributions. The first is the $\mathrm{sinc}^2(\Delta kL/2)$ phase-matching term, which determines the main spectral acceptance structure. A longer crystal length leads to a narrower phase-matching acceptance with respect to $\Delta k$, and therefore makes the calculated response more sensitive to wavelength-dependent phase mismatch.

The second contribution is the slowly varying wavelength-dependent factor$\left(n_\mathrm{s} n_\mathrm{p} n_\mathrm{up}\lambda_\mathrm{s}\lambda_\mathrm{up}\right)^{-1}$, which changes gradually across the SWIR range. This factor does not determine the phase-matching wavelengths, but it modifies the relative amplitudes of the calculated response peaks. Therefore, the peak-height asymmetry discussed below is included in the theoretical normalized quantum-efficiency spectra.

\subsection{Experimental setup and external quantum-efficiency measurement}

\begin{figure}[htbp]
\centering\includegraphics[width=\columnwidth]{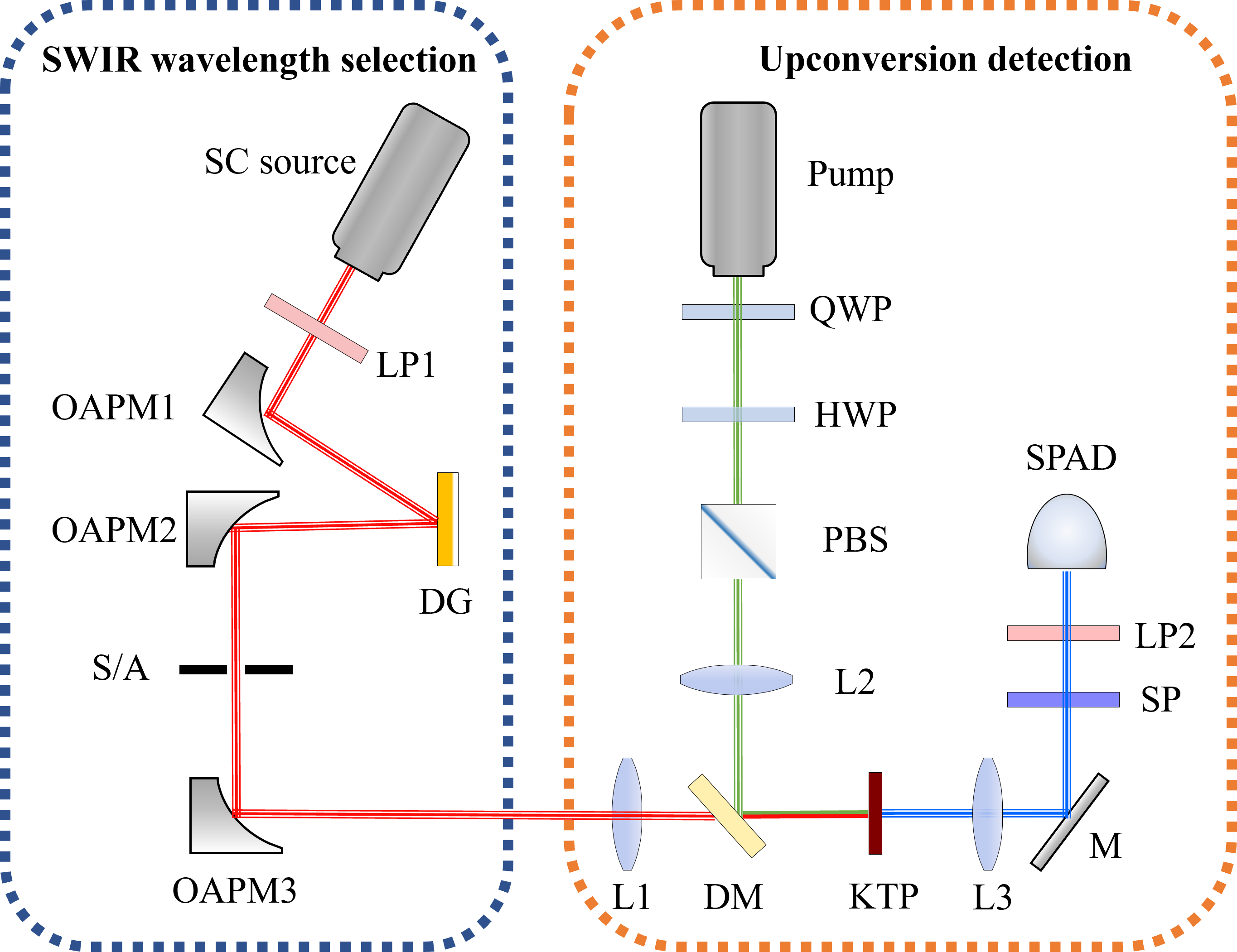}
\caption{
Schematic of the experimental setup. SC source, supercontinuum source; LP1, long-pass filters in the SWIR signal path; OAPM, off-axis parabolic mirror; DG, diffraction grating; S/A, slit or adjustable aperture; M, mirror; L, lens; DM, dichroic mirror; HWP, half-wave plate; QWP, quarter-wave plate; PBS, polarizing beam splitter; SP, short-pass filters; LP2, long-pass filter in the upconverted-light path; SPAD, single-photon avalanche diode.
}
\label{fig:1}
\end{figure}

Fig.~\ref{fig:1} shows a schematic of the experimental setup, which consists of two sections: SWIR wavelength selection and upconversion detection. The broadband SWIR light is spectrally selected to produce a narrowband signal, which is then converted to visible light through the NCPM process in the upconversion-detection module and subsequently detected.

The SWIR signal was derived from a pulsed supercontinuum source (430--2400 nm, repetition rate: 0.5 MHz, pulse duration: approximately 100 ps). Wavelengths below $1180~\mathrm{nm}$ were removed using a 1000 nm long-pass filter and a long-pass dichroic mirror (DMLP1180, Thorlabs), and the remaining broadband SWIR light was sent into the calibrated C--T monochromator described in Sec.~\ref{subsec:CTMODEL}. For measurements at wavelengths above 1600 nm, an additional 1500 nm long-pass filter was inserted into the signal path to improve the signal-to-noise ratio. 
The monochromator output was then focused by a lens and combined with the pump beam by a dichroic mirror.

The continuous-wave pump laser at $1017.15~\mathrm{nm}$ was generated by a tunable semiconductor seed laser (KT-TSL-1064, linewidth < 50 kHz, Anhui Kunteng Quantum
Technology), followed by a custom-built ytterbium-doped fiber
amplifier. After polarization control, the pump beam was focused with its waist positioned behind the KTP crystal, providing full spatial coverage of the SWIR signal spot at the crystal.

Three KTP crystals with lengths of $0.5~\mathrm{mm}$, $1.0~\mathrm{mm}$, and $2.0~\mathrm{mm}$ were compared. All crystals were operated at room temperature and had an aperture of approximately $6~\mathrm{mm}\times7~\mathrm{mm}$. Inside the KTP crystal, the SWIR signal and the $1017.15~\mathrm{nm}$ pump generated visible sum-frequency photons through the type-II NCPM interaction, where the pump, signal, and upconverted fields were polarized along the $y$, $z$, and $y$ axes of the crystal, respectively. 

After the crystal, two 850 nm short-pass filters were used to reject the residual pump and signal light, and a 550 nm long-pass filter was inserted to suppress the second harmonic of the pump and other short-wavelength noise. The filtered upconverted visible light was then coupled into a multimode optical fiber with a core diameter of 62.5 $\mu\mathrm{m}$ and detected by a silicon single-photon avalanche diode. During the comparative measurements, the three crystals were characterized using the same filtering, pump focusing condition, fiber-coupling, and Si-SPAD detection configuration. 

The external quantum efficiency was measured point by point at the calibrated signal wavelengths. At each wavelength, the signal power before the crystal, $P_s(\lambda_s)$, was first measured using a thermal power sensor (S401C, Thorlabs). During the power measurement, the background reading obtained without signal input was recorded and subtracted from the raw power. The background-corrected signal power at each wavelength was used to calculate the corresponding incident photon number, thereby correcting for wavelength-dependent variations in the input power. Upconversion measurements were then performed at the same wavelengths, and the net detected counts were recorded over an integration time of $T=10~\mathrm{s}$. The detailed procedure used to calculate the net counts is provided in the Supplementary Material.

The number of incident signal photons was calculated from the background-corrected signal power as
\begin{equation}
N_s=\frac{P_s(\lambda_s)T}{hc/\lambda_s},
\end{equation}
where $h$ is the Planck constant, $c$ is the speed of light in vacuum, and $T$ is the photon-counting integration time. The external quantum efficiency was then defined as the ratio between the background-corrected detected upconversion counts and the incident signal photon number:
\begin{equation}
\eta_{\mathrm{ext}}(\lambda_s)=\frac{N_{\mathrm{up}}}{N_s}.
\end{equation}

Finally, the obtained $\eta_{\mathrm{ext}}(\lambda_s)$ curves were normalized to their respective maximum values to obtain the normalized external upconversion response. 

\section{Results}
\subsection{Wavelength calibration and spectral-resolution verification}

\begin{figure}[ht]
\centering\includegraphics[width=\columnwidth]{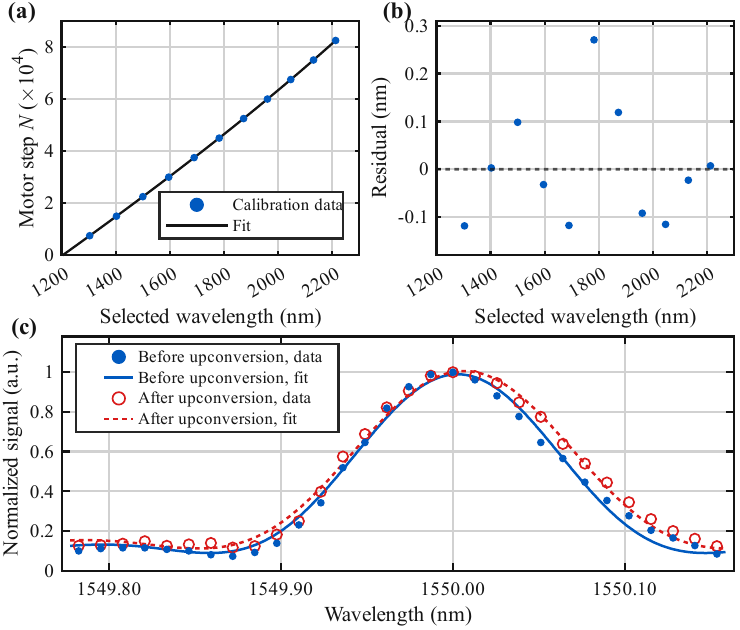}
\caption{Wavelength calibration and spectral-resolution verification of the C--T monochromator. 
(a) Calibration curve between the motor step number and the selected first-order SWIR wavelength, obtained using second-order visible diffraction. 
(b) Calibration residuals between the measured wavelength points and the fitted step--wavelength relation. 
(c) Spectral-resolution verification using a \(1550~\mathrm{nm}\) quasi-single-frequency laser. The output spectrum was measured both by direct power measurement before upconversion and by Si-SPAD photon-count detection after KTP upconversion.}
\label{fig:2}
\end{figure}
Fig.~\ref{fig:2} summarizes the wavelength calibration of the C--T monochromator and the spectral-resolution verification of the upconversion detection path.

Using the calibration method described in Sec.~\ref{subsec:CTMODEL}, the motor step number was mapped to the first-order SWIR wavelength. The calibrated wavelength points were fitted with the step--wavelength relation derived from the grating equation. 

To evaluate the calibration accuracy, the experimentally calibrated wavelengths were compared with the wavelengths predicted by the fitted curve. The residuals are shown in Fig.~\ref{fig:2}(b). The calibration residuals remained at the sub-nanometer level over the measured wavelength range, with a root-mean-square error of approximately $0.116~\mathrm{nm}$. This error is much smaller than the characteristic wavelength scales of the measured spectral features and therefore does not affect the comparison of the normalized response profiles of the KTP crystals.

The spectral-resolution verification was performed using a 1550 nm quasi-single-frequency laser. As shown in Fig.~\ref{fig:2}(c), the monochromator output was first measured directly with a power meter to obtain the normalized power data before upconversion. The same wavelength scan was then repeated using the upconversion detection path to obtain the normalized photon-count data after upconversion. The fitted curves give full widths at half maximum of 0.13~nm before upconversion and 0.14~nm after upconversion. The close agreement between these values indicates that the upconversion detection path introduces no obvious additional spectral broadening within the resolution of the present measurement.

\subsection{Length-dependent spectral responses of NCPM KTP crystals}

\begin{figure}[ht]
\centering\includegraphics[width=\columnwidth]{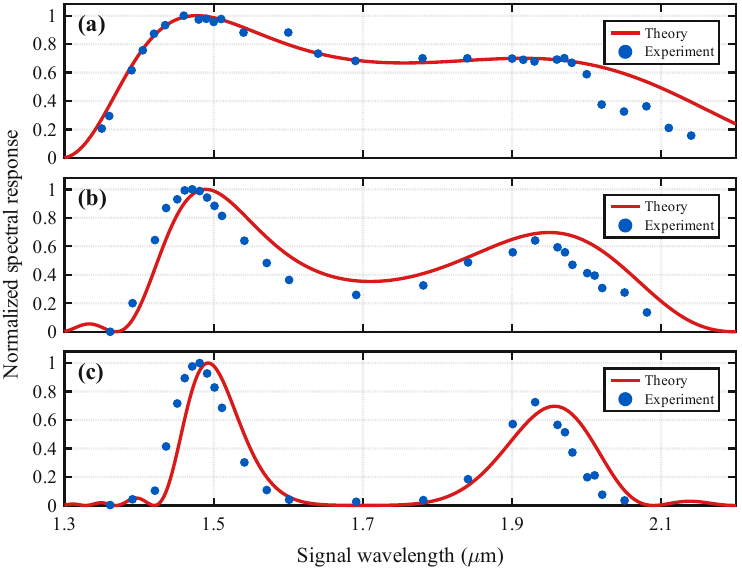}
\caption{Measured normalized external upconversion responses of NCPM KTP crystals with lengths of (a) 0.5 mm, (b) 1.0 mm, and (c) 2.0 mm, compared with the corresponding theoretical normalized quantum-efficiency spectra.
}
\label{fig:3}
\end{figure}

Fig.~\ref{fig:3} shows the length-dependent evolution of the measured and theoretical normalized spectral responses.
The three crystals exhibit clearly different spectral-response shapes. The 0.5 mm crystal shows a broad response over the measured SWIR range. Its long-wavelength maximum is broad and only weakly resolved, indicating that this short crystal provides broad spectral acceptance with only a shallow response valley between the two phase-matching maxima. In contrast, the 1.0 and 2.0 mm crystals exhibit progressively clearer double-peak structures. The response valley becomes more pronounced as the crystal length increases, reflecting the reduced tolerance to phase mismatch for a longer nonlinear interaction length.

\begin{figure}[h]
\centering\includegraphics[width=\columnwidth]{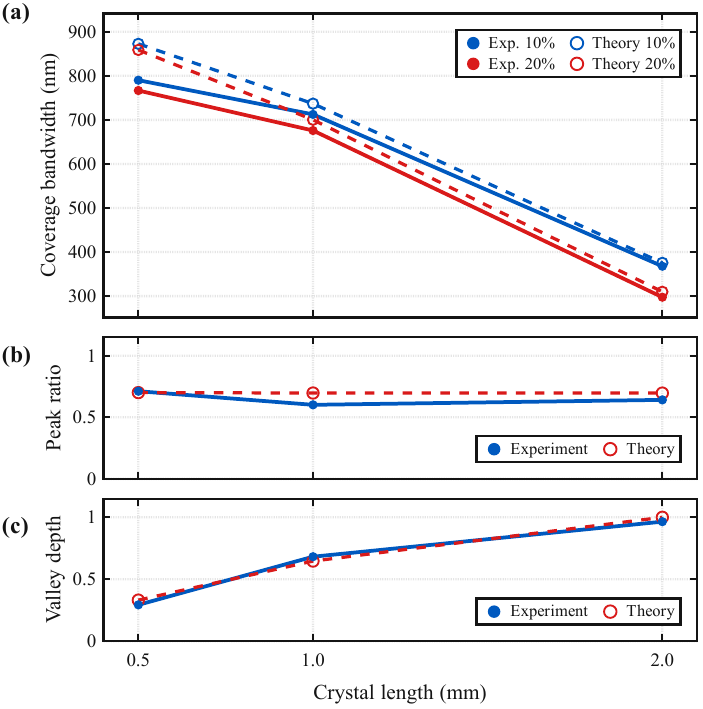}
\caption{
Extracted length-dependent spectral-response metrics of NCPM KTP crystals.
(a) 10\% and 20\% coverage bandwidths as functions of crystal length.
(b) Peak ratio, defined as the ratio of the long-wavelength peak response to the short-wavelength peak response.
(c) Valley depth, defined as $1-R_\mathrm{valley}$, where $R_\mathrm{valley}$ is the normalized response at the valley between the two phase-matching peaks.
Solid symbols represent experimental values extracted from the measured normalized external upconversion responses, and open symbols represent the corresponding theoretical values.
}
\label{fig:4}
\end{figure}

The extracted spectral-response metrics are plotted in Fig.~\ref{fig:4}. As the crystal length increases, both the 10\% and 20\% coverage bandwidths decrease monotonically. This trend is consistent with the narrower phase-matching acceptance of longer crystals.

In addition to the coverage bandwidths, the peak asymmetry and valley depth provide further information about the response shape. The theory predicts that the long-wavelength peak is approximately 70\% of the short-wavelength peak for the three crystal lengths. Experimentally, the 1.0 and 2.0 mm crystals show smaller peak ratios than the theoretical values, indicating an additional reduction on the long-wavelength side in the measured external response.

The measured valley depth increases from approximately 0.294 for the 0.5 mm crystal to 0.682 for the 1.0 mm crystal and 0.964 for the 2.0 mm crystal. The theoretical valley depths show the same length-dependent evolution.

\subsection{Representative pump-power dependence of the external quantum efficiency}
\label{subsec:pump_power_scaling}

\begin{figure}[htbp]
\centering
\includegraphics[width=\columnwidth]{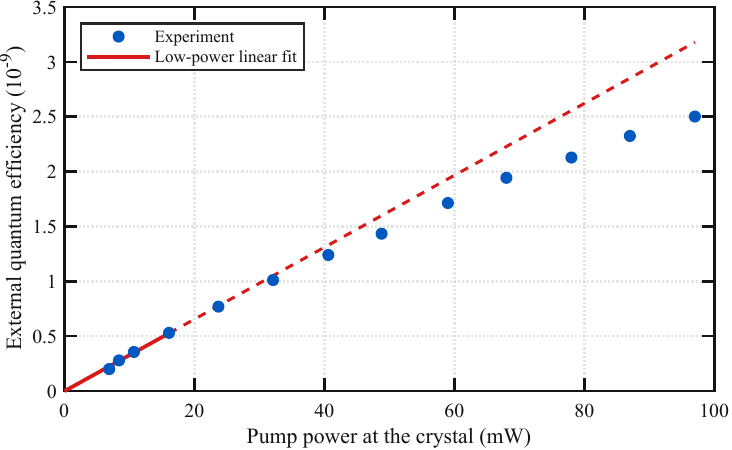}
\caption{Representative pump-power dependence of the external quantum efficiency measured near the 1480 nm short-wavelength response peak of the 2.0 mm NCPM KTP crystal. The solid line represents a zero-intercept linear fit to the four low-pump-power points from 6.9 to 16.1 mW, while the dashed line shows the extrapolation of this linear relation to higher pump powers.}
\label{fig:5}
\end{figure}

A representative pump-power dependence measurement was performed near the short-wavelength response peak of the 2.0 mm KTP crystal to estimate the system-level external quantum-efficiency slope. The signal wavelength was fixed near 1480 nm, and the signal power measured before the KTP crystal was maintained at $0.19~\mathrm{mW}$. The $1/e^2$ pump-beam diameter at the crystal was approximately $693~\mu\mathrm{m}$. The pump power was varied by adjusting the amplifier drive current and was independently measured before the crystal for each current setting. Because the amplifier output power depends nonlinearly on the drive current, the resulting pump-power points were not equally spaced. For each pump-power setting, the external quantum efficiency was calculated, with background subtraction and signal photon-number calculation performed as described above.

The solid line in Fig.~\ref{fig:5} is a zero-intercept linear fit to the four low-pump-power points (6.9--16.1 mW), yielding
\begin{equation}
\eta_\mathrm{ext}
=
\xi_\mathrm{ext} P_\mathrm{p},
\label{eq:external_slope_efficiency}
\end{equation}
where $P_\mathrm{p}$ is the pump power at the crystal and $\xi_\mathrm{ext}$ is the measured external quantum-efficiency slope. The fitted value was
\begin{equation}
\xi_\mathrm{ext}
=
3.28\times10^{-8}~\mathrm{W^{-1}}.
\label{eq:external_slope_efficiency_value}
\end{equation}
The dashed line extrapolates this fit to higher pump powers.

The measured external quantum efficiency continued to increase with pump power and reached a maximum recorded value of $2.50\times10^{-9}$ at $97~\mathrm{mW}$. The gradual departure from linearity at higher pump powers is likely attributable to count-rate compression of the Si-SPAD, because the detected photon-counting rate reached several million counts per second and entered its nonlinear response regime. The signal power was not reduced further because the limited sensitivity of the thermal power sensor prevented reliable power calibration at lower signal levels. Consequently, the high-pump-power measurements were acquired at count rates where the measured external quantum efficiency could be underestimated. The observed curvature is therefore not interpreted as unambiguous evidence of saturation of the nonlinear conversion process, and only the data in the low-pump-power region were used to extract $\xi_\mathrm{ext}$.

The obtained external quantum-efficiency slope is a system-level quantity that characterizes the present optical coupling, upconversion, collection, and detection configuration, rather than an intrinsic conversion coefficient of the KTP crystal alone.

\section{Discussion}
\label{sec:discussion}
The measured spectral responses agree with the calculated quantum-efficiency spectra. Shorter crystals provide broader coverage, whereas increasing the interaction length narrows the phase-matching acceptance and produces more distinct peaks.

Quantitative deviations nevertheless remain, particularly on the long-wavelength side for the 1.0 and 2.0 mm crystals. The measured quantity is a normalized external upconversion response, which includes not only the intrinsic phase-matching response of the KTP crystal but also wavelength-dependent system factors. Since the input SWIR power was measured before the crystal and corrected at each wavelength, the deviations are unlikely to originate from incident signal-power variations alone. Other wavelength-dependent factors that were not independently calibrated, including the SWIR transmission of the KTP crystals and the visible-light filtering, collection, fiber coupling, and detector response after upconversion, may also contribute to the measured external response. Therefore, the observed long-wavelength deviation likely reflects residual wavelength-dependent effects in the present measurement configuration.

The external quantum efficiency could be further improved by increasing the pump intensity and optimizing the detection configuration. In the present setup, the maximum usable pump power was constrained by the linear count-rate range of the Si-SPAD, and higher pump powers would require a detector with a wider dynamic range, such as a linear-mode silicon APD or a photomultiplier tube. The pump focusing condition was chosen to provide full spatial coverage of the SWIR signal spot at the crystal, which reduced the peak pump intensity. The measured efficiency therefore reflects a system-level trade-off among spatial overlap, pump intensity, and detector dynamic range.

\section{Conclusion}
We investigated the length-dependent SWIR upconversion spectral response of noncritically phase-matched KTP crystals using a calibrated C--T monochromator to measure the normalized external upconversion response at different wavelengths. The responses of 0.5, 1.0, and 2.0 mm KTP crystals were compared with theoretical normalized quantum-efficiency spectra calculated from the phase-matching model. The extracted coverage bandwidths, peak ratios, and valley depths show that the crystal length serves as an effective design parameter for shaping the external response.

A representative pump-power measurement yielded an external quantum-efficiency slope of $\xi_\mathrm{ext}=3.28\times10^{-8}~\mathrm{W^{-1}}$ in the low-pump-power region. This value characterizes the present detection configuration and provides a reference for future efficiency optimization. Further improvements in pump intensity, collection efficiency, detector dynamic range, and monochromator resolution would increase the sensitivity and spectral resolution of this SWIR upconversion detection scheme.

\begin{backmatter}
\bmsection{Funding}
We would like to acknowledge the support from the National Key Research and Development Program of China (2022YFB3903102, 2022YFB3607700), National Natural Science Foundation of China (NSFC)(62435018), Innovation Program for Quantum Science and Technology (2021ZD0301100), USTC Research Funds of the Double First-Class Initiative (YD2030002023), and Research Cooperation Fund of SAST, CASC (SAST2022-075).


\bmsection{Disclosures}
The authors declare no conflicts of interest.

\bmsection{Data Availability Statement}
Data underlying the results presented in this paper are not publicly available at this time but may be obtained from the authors upon reasonable request.

\end{backmatter}

\bibliography{main}

\end{document}